\documentclass[manuscript]{aastex}

\slugcomment{submitted}

\shorttitle{Feedback from massive stars and gas expulsion from proto-Globular Clusters}
\shortauthors{Calura et al.}

\begin{document}


\title{Feedback from massive stars and gas expulsion from proto-Globular Clusters}


\author{F. Calura\altaffilmark{1}, C. G. Few\altaffilmark{2}, D. Romano\altaffilmark{1} and A. D'Ercole\altaffilmark{1}} 
\affil{1: INAF, Osservatorio Astronomico di Bologna, via Ranzani 1, I-40127 Bologna, Italy}
\affil{2: School of Physics, University of Exeter, Stocker Road, Exeter EX4 4QL, UK}



\begin{abstract}
Globular clusters are considerably more complex structures than previously thought, harbouring at least two 
stellar generations which present clearly distinct chemical abundances. 
Scenarios explaining the abundance patterns in globular clusters mostly assume 
that originally the clusters had to be  much more massive than today, and that 
the second generation of 
stars originates from the gas shed by stars of the first generation (FG). 
The lack of metallicity spread in most globular clusters further 
requires that the supernova-enriched gas ejected by the FG is completely lost within $\sim~30$Myr, a
hypothesis never tested by means of three-dimensional hydrodynamic simulations. 
In this paper, we use 3D hydrodynamic simulations including stellar feedback from winds and supernovae, radiative cooling and 
self-gravity to study whether
a realistic distribution of OB associations in a massive proto-GC of initial mass $M_{tot} \sim 10^{7} M_{\odot}$ 
is sufficient to expel its entire gas content. 
Our numerical experiment shows that the coherence of different associations plays a fundamental role: as the bubbles 
interact, distort and merge, they carve narrow tunnels which reach deeper and deeper towards the innermost cluster regions, 
and through which the gas is able to escape.  
Our results indicate that after $3$ Myr, the  feedback from stellar winds is responsible for the removal of  $\sim$40$\%$ of the pristine gas, and that after $14$ Myr,  99$\%$ of the initial gas mass has been removed.
\end{abstract}

\section{Introduction}
Globular clusters (GCs) are now known to host at least two stellar
populations, distinct in chemical composition (e.g. Carretta et
al. 2009a,b). The different populations differ mostly in light element
(C, N, O, Na, Mg, Al) abundances, while star-to-star abundance
variations in elements heavier than Si are generally not detected (Gratton et al. 2004).\\
So far, various models have been put forward to explain the abundance
pattern observed in GCs and to understand the origin of their multiple
populations.  The most popular models consider a scenario in which a
second generation (SG) of stars forms from the gas ejected by either
first generation (FG) asymptotic giant branch (AGB) (Cottrell \& Da
Costa 1981; D'Antona \& Caloi 2004; Karakas et
al. 2006; D'Ercole et al. 2008; Bekki 2011), or FG fast rotating
massive stars (Prantzos \& Charbonnel 2006; Decressin et al. 2007).  In both scenarios, the GC
precursors have to be more massive (by factors between 5 and 20) than
the objects we see today. In the AGB scenario, a key
 question is whether the gas enenergized by the FG of supernovae
 (SNe) is lost by the system at the time the SG forms. 
 This scenario requires that the pristine gas has to
 be completely lost if very O-poor SG stars may form as observed in
 some GCs (Carretta et al. 2009a; D'Ercole et al. 2010). 
Therefore, understanding whether after the formation of the FG the residual gas is expelled
is fundamental in order to support this scenario. 
 
Previous studies indicate that in clusters of mass $\sim 10^7\,M_{\odot}$, 
once radiative losses are taken into account, 
the combined energy of all SNe can be lower than the potential energy of the gas
(Dopita \& Smith 1986; Baumgardt et al. 2008), hence 
a significant fraction of the initial gas will be retained.
However, there has been no attempt to study gas loss from such a massive 
proto-GC by means of realistic three dimensional (3D) simulations. 
The use of 3D hydrodynamic simulations is crucial in order to investigate in detail the structure of 
the outflow and how the gas can escape from the system, 
as well as the spatial extent of such potential escape routes. 
Moreover, taking into account a realistic 
distribution of stars in a cluster is crucial, in order to assess the relative, contrasting roles of 
stellar feedback and radiative losses. \\
In this work, we run 3D Adaptive Mesh Refinement (AMR) hydrodynamic simulations 
to investigate the response 
of the typical gas-rich GC precursor of mass $10^7\,M_{\odot}$ 
to the action of multiple feedback sources, 
including both stellar winds and supernovae, occurring in stellar associations. 
Our aim is to assess on firmer grounds whether the system can lose its gas via internal processes alone. 
This Letter is primarily aimed at presenting our results on the main subject of our investigation, i.e. 
whether the feedback of stellar winds and SNe in one massive proto-cluster is sufficient to remove its gas. 
A more extended description of our results is postponed to a future paper.

\section{Simulation setup}
\label{set}
The simulations have been performed using the RAMSES code (Teyssier 2002). 
The GC is initially composed of a self-gravitating gas distribution in hydrostatic equilibrium, 
and of a FG of stars already in place.
To study the impact of the feedback from massive stars in a massive precursor of nowadays GCs, 
we assume a total initial mass $M_{tot} \sim 10^{7} M_{\odot}$, of which 
$M_{*}=3 \times 10^{6} M_{\odot}$ are in stars. 
This equates to a star formation efficiency (SFE) $\sim~0.4$, 
consistent with the general requirement that the formation of 
GCs occurs preferentially in high-pressure regions, characterised by SFEs somewhere between 0.25 and 0.5 
(e.g. Elmegreen \& Efremov 1997). 

The initial, non-rotating 
mass distribution follows a Plummer (1911) density profile, 
characterised by a density $\rho(r)$ expressed as:
\begin{equation}
\rho(r) = \frac{3M_{tot}}{4\pi\, a^3} \left(1+\frac{r^2}{a^2}\right)^{-\frac{5}{2}}.
\label{plum}
\end{equation}
In this equation, $r$ is the radius and $a=27$~pc is the Plummer radius\footnote{This value is chosen following 
the prescriptions of D'Ercole et al. (2008), who use for their cluster model a King (1962) 
density profile with core radius $r_c=6.3$~pc and truncation radius $r_t=200$~pc. Here, we use instead a Plummer 
profile with the same half-mass radius, i.e. $r_{half} = 1.3 a = \sqrt{r_c \, r_t}=35$~pc. }. This corresponds to a rather dense system, 
characterised by a central gas density of $5 \times 10^3$~cm$^{-3}$.\\
A static gravitational potential is assumed for the stellar component, expressed by:
\begin{equation}
\Phi_*(r) = -\frac{G M_{*}}{\sqrt{r^2 + a^2}},  
\end{equation}
where $G$ is the universal gravitational constant. 
The self-gravity of the gas is taken into account and the potential due to the gas distribution $\Phi_{gas}$ is computed  
at each timestep by solving the Poisson equation in an adaptive grid framework, as described in Teyssier (2002). 

For the gas component, the initial pressure profile is determined by solving the hydrostatic equilibrium equation:
\begin{equation}
\frac{dP}{dr}= \frac{-G\, M_{tot}(r)\,\rho(r)}{r^2}. 
\end{equation} 

In our simulation, the computational box has a volume of $L_{box}^3 = (162$~pc$)^3$ and is characterised by a 
maximum resolution of $L_{box}\,0.5^{levelmax=8}\sim\,0.6$~pc. 
This assumption, together with the minimum temperature floor of  $T_{min}=5000~K$ (see Sect.~\ref{feed}) assures that the Jeans length is always well above 4 times the maximum resolution, 
a condition sufficient to avoid artificial fragmentation in simulations of self-gravitating gas (Truelove et al. 1997).

Free outflow boundary conditions are used. 
The refinement stategy is both geometry- and discontinuity-based. 
At each timestep, in correspondence of the position of each source 
a number of cells at the highest refinement level were created as described in Sect.~\ref{feed}. 

A passive tracer was created to study the evolution of the metallicity $Z$ within each cell, with 
$Z = \frac{\rho_Z}{\rho_{gas}}$, where $\rho_Z$ is the density of metals in a given cell. 
For the initial metallicity, we have chosen the value $Z_{ini} = 0.001 Z_{\odot}$ (D'Ercole et al. 2008).

\subsection{Heating and cooling prescriptions}
\label{feed}
The heating sources are the massive stars (MS,~$M>8~M_\odot$), which are allowed to release mass and energy in 
both their pre-supernova (pre-SN) and supernova (SN) phases. 
Adopting a standard IMF (e.g. Kroupa 2001), 
$\sim$0.01 MS are born per $M_\odot$ of stars formed.

The total MS population has been split into a number of separate OB associations (OBAs); 
each association is allowed to continuously inject energy and mass for 30 Myr, 
a time which roughly corresponds to the lifetimes of stars of $M=8 M_{\odot}$. 

The energy and mass input rate from OBAs is approximated as continuous (Mac Low \& McCray 1988) and constant in time. 
During the pre-SN (i.e. at times  $t \le 3 $ Myr) and SN phases, 
we assume that an OBA which includes $N$ massive stars ejects $N\,~\dot{m}=N\,5\times 10^{-8}$ and  $N\,\dot{m}=N\,~4~\times10^{-7}$~$M_{\odot}/yr$ and 
emits $N\,~l=N\,~3\times~10^{35}$ and  $N\,~l=N\,~7\times10^{35}\,erg/s$, respectively (Leitherer et al. 2014\footnote{The mass and energy injection rates used here 
are derived from the Starburst99 prescriptions (Leitherer et al. 2014) and 
calculated for a simple stellar population of sub-solar metallicity.})

The MS number $N$ is sampled randomly 
from the power-law distribution 
\begin{equation}
f(N) \propto N^{\alpha}, 
\end{equation}
with $\alpha=-2$ and $30 \le N \le 2000$ (Higdon \& Lingenfelter 2005; Melioli et al. 2009). 

The OBAs have been scattered in the simulation volume 
by sampling randomly from the mass distribution expressed by  Eq. ~\ref{plum}. 
The probability of having an OBA located at a distance $r_j$ from 
the center of the computational box is given by the number 
\begin{equation}
X_j = \frac{\int_0^{r_j} \rho(r) dV}{\int_0^R \rho(r) dV}, 
\end{equation}
where $R=L_{box}/2$ and $X_j$ is a random number, with $0\le X_j \le1$. 
Standard coordinate transformations are used to derive  cartesian coordinates of each OBA, 
and the values for the quantity $sin\,\theta$ and the azimuthal angle $\phi$ were determined by sampling randomly from 
flat distributions. 
To avoid the occurrence of diamond-shaped shock fronts we give our energy sources a radial size of $\Delta R=2$~pc ($\sim$3 times the maximum resolution).

The total thermal energy and mass injected into the ISM per unit volume by the OBA in the time step $\Delta \, t$ are $\Delta E= \dot{E}_* \Delta\,t$ 
and $\Delta M=\dot{\rho}_* \Delta\,t$, respectively, where $\dot{\rho}_*=\frac{\dot{m}N}{V}$, $\dot{E}_*=\frac{N\,l}{V}$ and 
$V=\frac{4}{3}\pi\Delta R^3$ is the volume of a single OBA. \\

The cooling function implemented in RAMSES takes into account both atomic (i.e., due to H and He) 
and metal cooling, due to the presence of metals (see Few et al. 2014). 
The contribution from metals at temperatures above $10^4$ K is accounted for through a
fit of the difference between the cooling rates calculated at solar metallicity and those at zero
metallicity using the photoionisation code
CLOUDY (Ferland et al. 1998). At lower temperatures, metal fine-structure cooling rates are taken from Rosen \& Bregman (1995). 

For computational reasons, a constant-value temperature floor $T_{min}=5000 K$ is assumed.

In general, in high density regions the feedback energy returned to the ISM is radiated away extremely quickly (Katz 1992); 
so far, various methods have been proposed to prevent immediate radiative loss of the energy injected by massive stars (Thacker \& Couchman 2000). 
In this work, we switch off cooling in suitable grids following the prescriptions of Teyssier et al. (2013). 
We assume that each OBA injects in the ISM an amount of non-thermal energy 
of the same order of magnitude as the thermal energy. 
In each cell $(i,j,k)$ and at the time $n$, we introduce a non-thermal velocity dispersion $\sigma_{turb}$, defined as:
\begin{equation}
\sigma_{turb} = \sqrt{ 2\frac{\epsilon_{turb,i,j,k}}{\rho_{i,j,k}^n} }, 
\end{equation}
where the passive tracer $\epsilon_{turb,i,j,k}$ is the cumulative energy density injected by the OBAs and 
$\rho_{i,j,k}^n$ is the density in the cell at the $n-$th timestep. 
As assumed in Teyssier et al. (2013), we switch off cooling in any cell for which $\sigma_{turb}>10 \, km/s$.
This simple prescription is aimed at capturing roughly the sub-grid non-thermal processes, 
such as turbulence, which allow a more efficient coupling of the energy associated with stellar 
feedback to the gas component (see Teyssier et al. 2013).

\begin{figure}
\epsscale{.65}
\plotone{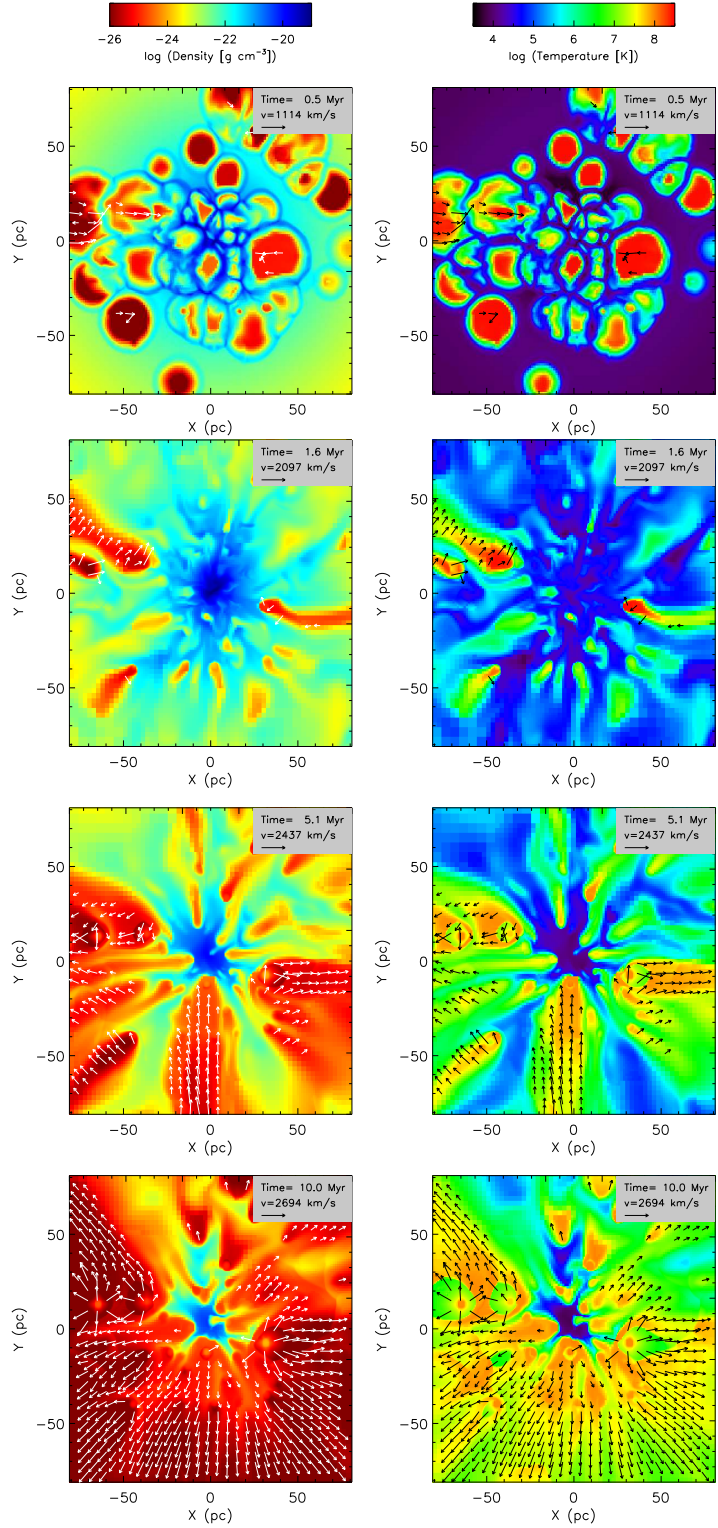}
\caption{Two-dimensional density (left panels) and temperature (right panels) maps computed for 
our highest resolution simulation 
at various evolutionary times. The arrows represent the velocity field (see text for details). 
\label{fig2}}
\end{figure}

\begin{figure}
\epsscale{.990}
\plotone{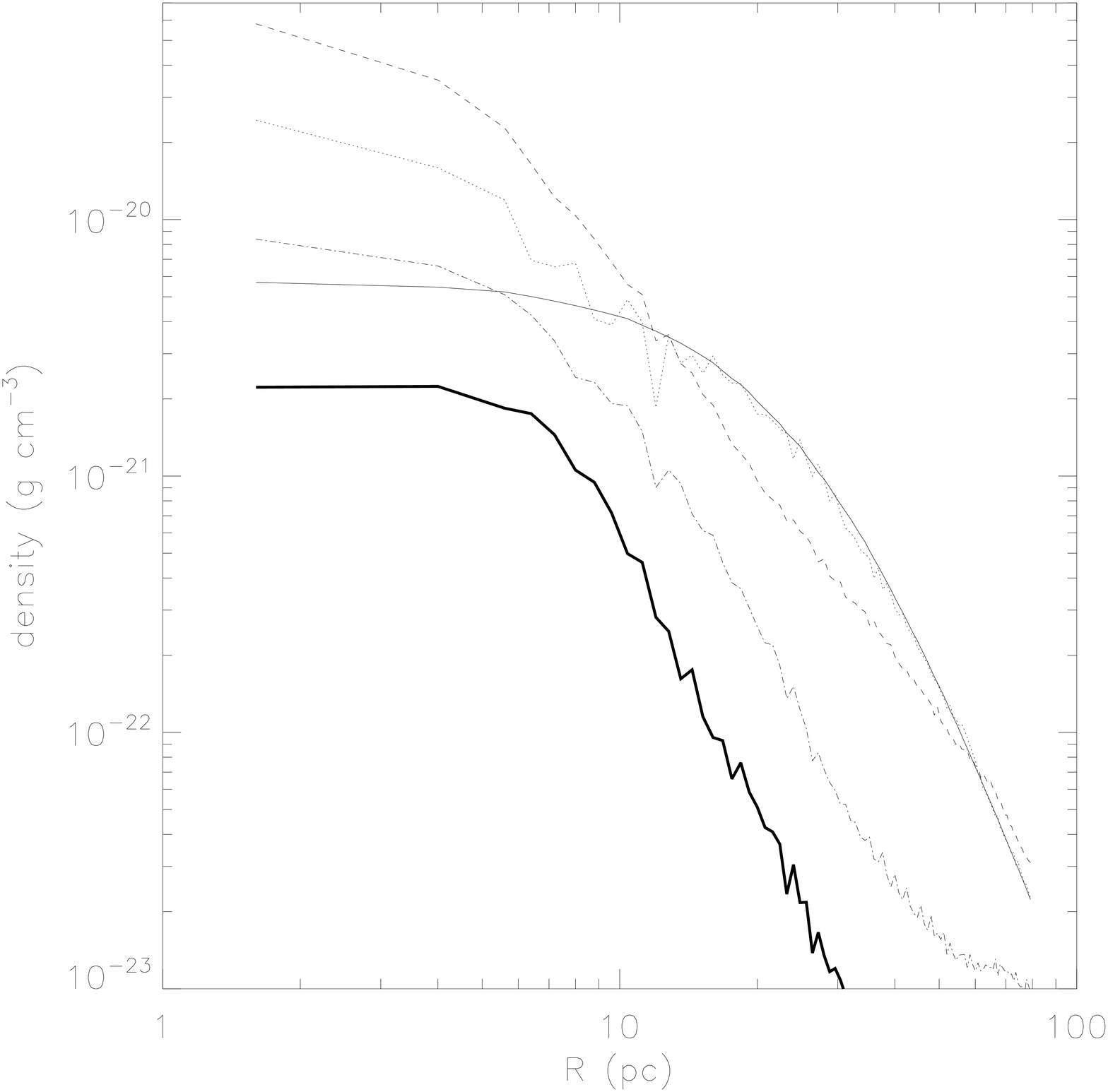}
\caption{Radial density profiles 
calculated for our highest resolution simulation at $t=0$~Myr (thin solid line), $t=0.5$~Myr (dotted line), $t=1.6$~Myr (dashed line), 
$t=5.1$~Myr (dash-dotted line) and $t=10$~Myr (thick solid line). 
\label{fig3}}
\end{figure}

\section{Results}
\subsection{Hydrodynamic evolution of the wind- and SN-driven gas}
\label{sec_map}
The complex structure of the interstellar medium as driven by stellar wind activity and SNe is visible in 
Figure~\ref{fig2}, where we show 2D density and temperature maps calculated for 
our GC model in the x-y plane at various simulation times. 
The velocity fields are also shown in Fig.~\ref{fig2} as arrows, 
drawn only for velocities with values greater than $\sim 0.02$ of the maximum value, 
which is reported in the top-right inset of each panel, along with the simulation time. 
As shown in Fig.~\ref{fig2}, several cavities filled
with hot, low-density gas created by the activity of the stellar winds are already visible at 0.5 Myr, when 
a few superbubbles expanding supersonically in the unperturbed 
medium still conserve a nearly round shape. 
This is true mainly for the isolated ones, located in low-density regions. 
On the other hand, bubbles originating from sources with one or more OBAs nearby 
show significant distortions and appear asymmetrically compressed even at early times. 
After the merging of multiple blowouts, the shells around each bubble have dissolved 
and a high-density, low-temperature region is visible in the center of the cluster at $t=1.6$~Myr, 
created by the coupled action of multiple wind-blown bubbles, 
pushing material in opposite directions, and by the gravity of the system.  
Outside the central region, at 1.6 Myr the low density gas 
occupies elongated shapes which extend out to the boundaries. 

The gas reaches the highest velocities within these lowest density regions; as visible 
at the bottom of the panels computed at 5.1 Myr, some of the velocities are still 
directed preferentially towards the centre, i.e. they move mostly under the influence of gravity. 
In the meantime, the extent of the high-density central region is reducing and the 
filling factor of the hottest, lowest density cavities is increasing.
Some of the hot gas is channelling towards the boundaries, as shown by the  outwards-directed velocity field, 
in particular at the left and right sides of the map at 5.1 Myr. 

The maps calculated at 10 Myr show how nearly all the space is now filled with low-density, high-temperature gas. 
In nearly every point of the plot, the velocities are now directed outwards. 
The central cold filament occupies a limited portion of the plane, without affecting substantially the large-scale 
motion of the gas. 

\subsection{Evolution of the radial profiles}
\label{sec_prof}
\begin{figure}
\epsscale{.990}
\plotone{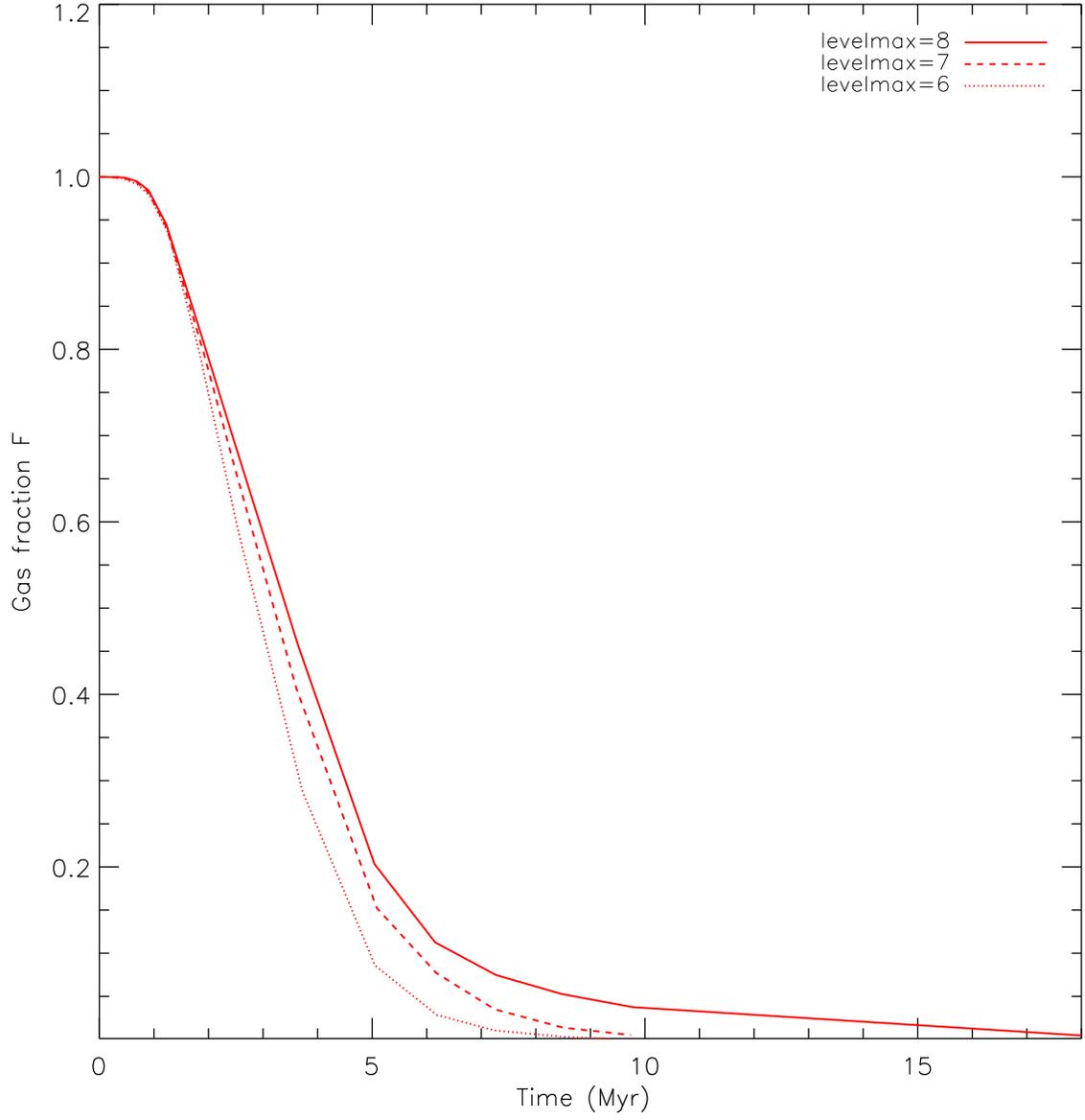}
\caption{Time evolution of the gas fraction, normalised to the initial gas mass, calculated at different resolutions, indicated by the maximum 
refinement level as reported in the top-right corner. 
\label{fig4}}
\end{figure}

In Fig.~\ref{fig3}, we show the time evolution 
of the mass-weighted density profiles. 
The radial profiles reported in Fig.~\ref{fig3} have been calculated at 0, 0.5, 1.6, 5.1 and 10 Myr. 
At $t=0.5$~Myr, a significant increase in the central density is clearly visible. 
At 1.6 Myr, a considerable mass re-distribution has occurred, 
as traced by the density profile which in in the innermost 15 pc
has significantly steepened with respect to the initial one, with a density still increasing in the center and 
decreasing in the outskirts. Part of this mass re-distribution is due to the feedback from OBAs,  whose impact is 
already appreciable as they cause the removal of a non-negligible amount of the initial gas mass (see also Fig.~\ref{fig4}). 

As explained in Sect.~\ref{feed}, at 3 Myr SNe take over as primary feedback sources, 
with mechanical luminosities significantly larger than those characterising Pre-SN. 
As a consequence of this, between $t=1.6$~Myr and $t=5.1$~Myr the system undergoes significant removal of the gas. 
The radial profile calculated at 10 Myr is considerably lower than the initial one. At this time, more than 
$\sim 90 \% $ of the initial gas mass has been removed. 
Our results thus confirm the assumption of the AGB scenario for the multi-population GCs, i.e. that the energy restored by 
massive stars is sufficient to clear the system of all the pristine gas.

\subsection{Discussion}
Our numerical experiments show that the simultaneous action of several feedback sources realistically 
distributed within a GC leads to the removal of the gas initially present in the system. 
In a configuration where the feedback sources are scattered in the simulation volume as described in Sect.~\ref{feed}, 
the character of the outflow is not as stationary and smooth as described in works assuming a more uniform source distribution in 
starbursting and dwarf galaxies (Chevalier \& Clegg 1985, Silich \& Tenorio-Tagle 1998). 
Even if the energy and mass inputs of the OBAs are continuous, no homogeneous outflow is generated by the OBAs, 
but rather an inhomogeneous flow through which the gas leaks out along the paths of least resistance. 
The resistance is created by the dense filaments as illustrated by the velocity field, in that 
the fast, low-density gas is often deviated whenever it encounters an extended, high-density sheet. 
Such porous and filamentary structures are common in 3D hydrodynamic 
simulations of starburst-driven winds (Rodriguez-Gonzalez et al. 2008; Cooper et al. 2008; Melioli et al. 2013).  

A major role in the generation of the outflow is played by the feedback sources during the pre-SN phase, i.e. when the feedback is 
dominated by stellar winds, active within the first 3 Myr of evolution. 
These sources are not only important since they can carve large, hot and tenuous cavities around stellar associations 
(see Breitschwerdt \& de Avillez 2006; Hopkins et al. 2012a; Rogers \& Pittard 2013; Dale et al. 2014, Geen et al. 2015), 
in which SNe explode and can restore energy at high thermalization efficiencies, but also as highly efficient feedback agents (Leitherer et al. 1992; Rosen et al. 2014). 

The coherence of different associations is a fundamental ingredient: as the bubbles 
interact, distort and merge, the narrow tunnels reach deeper and deeper towards the innermost cluster regions. 
The idea that the coherence of different sources is a basic requirement for a large thermalization efficiency 
is not new: in fact, previous works have already shown that single, episodic, isolated feedback sources are known to be 
less efficient than multiple sources acting simultaneously (Nath \&  Shchekinov 2013; Sharma et al. 2014). 
In our case, different OB associations act coherently and are able to generate expanding bubbles which rapidly 
interact and merge. 

Our simulations indicate that a timescale of a few Myr is sufficient for a significant amount of 
gas depletion to occur within GCs. We find that after $14$ Myr,  99$\%$ of the initial gas mass has been removed.

This is clearly visible in Fig.~\ref{fig4}, where we show the time evolution of the gas fraction, normalized to the initial value, 
computed at different resolutions.
Interestingly, current observational studies of very young,  massive clusters (M$\ge10^6M_{\odot}$)
show that they are able to clear out their natal gas within a few Myr after their formation (e.g., Bastian et al. 2014), 
although the mechanisms responsible for such a rapid gas removal of gas is not entirely known. 
The most plausible candidates could be massive stars in the stellar wind phase, although 
the observed bubble sizes appear to be larger than those predicted by hydrodynamic models including feedback from  stellar winds (Dale et al. 2014). 
In the future, it will be important to study, with simulations like the ones presented here, the causes of this discrepancy, 
in particular the role of the assumed initial cluster compactness and possibly the impact of various feedback processes, such as radiation pressure.\\

A look at Fig.~\ref{fig4} suggests also that the mass decrease is rather robust with respect to resolution, even if 
in the simulations performed at lower resolution, gas depletion seems to occur slightly faster. 
In general, numerical convergence with AMR simulations is more difficult to obtain than with uniform grid codes (Li 2010; Schmidt et al. 2014). 
In our case, 
the differences at various resolutions could be partially ascribed to the Poisson solver, 
which sometimes fails to converge at lower resolution. 

We have considered a cluster of large initial mass ($M_{tot}\sim 10^7\,M_{\odot}$). This choice was motivated by the fact that 
according to current models of GC formation, 
at the moment of their formation, clusters had to be 5-20 times more massive than their present-day mass value (D'Ercole et al. 2008). 
This mass value is also a critical one for gas removal, as previous studies 
indicate that in systems of $10^7\,M_{\odot}$, SNe are unable to unbind all the natal gas
(Dopita \& Smith 1986; Baumgardt et al. 2008; Krause et al. 2012; Leigh et al. 2013). 
In principle, in lower mass clusters the gas removal should be more rapid, as densities and radiative losses should be lower and 
OB stars eject enough energy to disperse gas clouds within a few Myr (Baumgardt et al. 2008). 
As already mentioned, an important parameter to study in the future will be the cluster compactness, also because 
gas removal could in principle be more difficult in clusters with larger central densities. 
Furthermore, the role of mass segregation needs to be studied in detail. 
These topics are currently under investigation and results will be presented in a forthcoming work.

We conclude  with a few notes on the simplifications made in this work and on some neglected physical processes. 
HII photoionisation has been neglected and could produce important effects at such high densities (Hopkins et al. 2012a). 
To properly account for this process, 
full radiative transfer calculations would be required (see Hopkins et al. 2012a, 2012b). 
In principle, this process could further facilitate  gas removal from proto-GCs.

Star formation is assumed to take place instantaneously at the start of the run 
and, in principle, it could take place in the densest and coolest cluster regions. 
To study the effect of star formation longer timescales need to be considered, and the inclusion of this process has 
certainly a high priority in future studies. 
Magnetic fields have also been neglected; little is currently known about how they couple with stellar feedback and how this affects 
the evolution of the ISM. 
What is known so far is that magnetic fields could represent a further feedback agent. 
In fact, current studies of magnetic fields in SN remnants suggest that they 
could play a role in reheating the gas at epochs of the order of a few Myr (Balsara et al. 2008). 

Thermal conduction is also neglected in our simulation, however,  
numerical diffusion simulates this process originating spurious radiative losses which 
otherwise would be absent. Detailed calculations show that 
the amount of radiation lost due to diffusion 
is larger than that of a realistic heat conduction front (Recchi et al. 2001). 
A proper description of conductive fronts requires a very high resolution and would 
be extremely computationally demanding (Recchi \& Hensler 2007).

\acknowledgments
An anonymous referee is acknowledged for useful comments and 
C. Nipoti, P. Londrillo and F. Brighenti 
for several interesting discussions. 
Simulations were carried on at the CINECA center (Bologna, Italy) 
with CPU time assigned under ISCRA grants. 
FC and DR acknowledge Financial support from
PRIN MIUR 2010-2011, project ‘The Chemical and Dynamical
Evolution of the Milky Way and Local Group Galaxies’, prot.
2010LY5N2T and from INAF under the contract PRIN-INAF-2012. 
CGF acknowledges funding from the European Research Council for the
FP7 ERC starting grant project LOCALSTAR.
ADE acknowledges support from PRIN INAF 2014.

\end{document}